\documentclass[pre,twocolumn,superscriptaddress]{revtex4}

\usepackage{amsmath}
\usepackage{epsfig}

\begin{document}

\title{Approach to asymptotic behavior in the dynamics of the trapping reaction
}
\date{\today}
\author{Lucian~Anton}
\affiliation{Department of Physics and Astronomy,
University of Manchester, M13 9PL, U.K.}
\affiliation{Institute of Atomic
Physics, INFLPR, Lab 22, PO Box MG-36 R76900, Bucharest, Romania}
\author{Alan~J.~Bray}
\affiliation{Department of Physics and Astronomy,
University of Manchester, M13 9PL, U.K.}



\begin{abstract}

We consider  the trapping reaction  $A + B  \to B$ in  space dimension
$d=1$, where the $A$ and $B$ particles have diffusion constants $D_A$,
$D_B$ respectively. We calculate the probability, $Q(t)$, that a given
$A$ particle  has not  yet reacted at  time $t$.  Exploiting  a recent
formulation in which the $B$ particles are eliminated from the problem
we find,  for $t  \to \infty$, $Q(t)  \sim \exp[-(4/\sqrt{\pi})(\rho^2
D_Bt)^{1/2} - (C  \rho^2 D_A t)^{1/3} + \cdots]$,  where $\rho$ is the
density of $B$ particles and $C \propto D_A/D_B$ for $D_A/D_B \ll 1$.

\end{abstract}

\maketitle


\section{Introduction}
\label{Introduction}

The trapping reaction, $A + B \to B$, where $A$ and $B$ are diffusing
particles with diffusion constants $D_A$ and $D_B$ respectively, has
seen a resurgence of interest in the last few years \cite{MG,BB,BB2,
BMB,OBCM,MOBC}.  The problem is related to the two-species
annihilation reaction \cite{TW}, $A + B \to 0$, in the case where the
initial particle densities satisfy $\rho_A(0) < \rho_B(0)$.  In the
long-time limit $\rho_A(t) \ll \rho_B(t)$, and to study the asymptotic
decay of the $A$-particle density it suffices to study the survival
probability, $Q(t)$, of a single $A$ particle diffusing in a sea of
$B$ particles. In this limit, the two-species annihilation reaction is
equivalent to the trapping reaction. As the initial condition, the
$B$-particles are taken to be randomly distributed in space with
uniform density $\rho$, i.e.\ the probability to find a $B$-particle
in a volume element $dV$ is $\rho dV$.

The problem of determining the asymptotic behavior of $Q(t)$ has only
recently been solved \cite{BB,BB2}. The solution is limited to space
dimensions $d \le 2$. The asymptotic {\em form} of the solution has
been known for some time \cite{BL}. The $A$-particle survival
probability is given by
\begin{equation}
Q(t) \sim \left\{\begin{array}{ll}\exp(-\lambda_d t^{d/2}), & \ \ d<2 \\
\exp(-\lambda_2 t/\ln t), & \ \ d=2 \\
\exp(-\lambda_d t), & \ \ d>2 \\ \end{array} \right.
\label{BL}
\end{equation}
but the values of the constants $\lambda_d$ were not known. In refs.\
\cite{BB,BB2}, these were determined for $d \le 2$ by deriving upper
and lower bounds on $Q(t)$ and showing that they pinch as $t \to
\infty$.  The restriction to $d \le 2$ is a consequence of the
recurrence properties of random walks. For $d<2$, all particles can be
treated as zero-size points.  As a consequence, the combinations $\rho
(D_Bt)^{d/2}$ and $D_A/D_B$ are the only dimensionless quantities in
the theory.  A key result of \cite{BB,BB2} is that the coefficients
$\lambda_d$ in Eq.\ (\ref{BL}) are {\em independent of $D_A$} for $d
\le 2$, and therefore have the form $\lambda_d = a_d\,\rho D_B^{d/2}$,
where the quantities $a_d$ are pure numbers. The result of
\cite{BB,BB2} is
\begin{equation}
a_d =  \left\{\begin{array}{ll}(4\pi)^{d/2}\,(2/\pi d)\sin(\pi d/2), &
\ \ d<2 \\ 4\pi, & \ \ d=2 \\ \end{array} \right.
\label{a}
\end{equation}
For $d=2$, the $A$-particle has to  be given a non-zero size, but this 
size only enters as a cut-off in the logarithm in Eq.\ (\ref{BL}).

Although these results are asymptotically exact, numerical studies
\cite{MG,BB2} show that the asymptotic regime is not reached, either
in $d=1$ or $d=2$, after any time accessible to simulation, even using
sophisticated simulation methods \cite{MG} that allow values of $Q(t)$
as small as $10^{-70}$ to be achieved. It is natural, therefore, to
try to calculate the leading correction to the asymptotic
behavior. This is the principal aim of this paper. It should be stated
at the outset that this is a very challenging problem, and we have
obtained concrete results only in the limit $D_A \ll D_B$.

The starting point of our approach is an equation, first presented in
\cite{BMB}, that implicitly determines the survival probability of the
$A$-particle {\em for a given $A$-particle trajectory}, averaged over
all the $B$-particle trajectories and $B$-particle initial
conditions. In this equation, therefore, the $B$-particles have been
{\em eliminated from the problem}.  Their presence is felt only
through their density, $\rho$, and their diffusion constant, $D_B$,
which enters the equation through the $B$-particle diffusion
propagator. The final step in the calculation is to average the
survival probability over the $A$-particle trajectories, weighted with
the Wiener measure.

In the present  work we focus on computing  the leading pre-asymptotic
corrections  to  the known  asymptotic  behavior.   For simplicity  we
specialize to dimension  $d=1$, where the best numerical results are 
available. Including the  new correction, the  result  we obtain  is  
fully consistent  with  the available data.

The paper is organized as follows.  In section II we set up the
necessary formalism, eliminating the $B$ particles from the problem.
In order to make the paper self-contained, in section III we briefly
review the calculation of the leading asymptotic behavior.  The
pre-asymptotic corrections are discussed in section IV, at leading
nontrivial order in the ratio $D_A/D_B$. In the short Section V we
discuss how one might go beyond this leading order, and propose a
general form for the pre-asymptotic correction that interpolates
between small and large $D_A/D_B$. The paper concludes with a brief
discussion and summary of the results.

\section{Eliminating the $B$-particles}
\label{Elimination}

The  key step  in deriving  the `fundamental  equation' from  which we
start is a conceptual one.  We treat the $A$ and $B$ particles {\em as
if they were non-interacting},  and exploit the initial condition that
each  $B$  particle  is  randomly  located  anywhere  in  the  system.
Consider, for simplicity, a system of finite length, $L$, containing 
$N=\rho L$ $B$-particles (diffusion constant  $D_B$), randomly distributed 
within it, and  a single $A$  particle (diffusion constant  $D_A$), initially
located  at  the  origin.   Let  $z(t)$  be  the  $A$-particle's
trajectory, and  let $P(x,t)$ be the probability  that a given
$B$ particle, starting at $x$, has met the $A$ particle before
time  $t$. The  average of  this quantity  over the  initial position,
$x$, is $(1/L)\int dx\,P(x,t) = R(t)/L $, where $R(t)$
is an  implicit functional of $z(t)$.  The  probability that $n$
distinct $B$ particles have met  the $A$ particle, averaged over their
initial    positions,    is    $p_n(t)    =    \binom{N}{n}    (R/L)^n
(1-R/L)^{N-n}$. Taking the limit $N  \to \infty$, $L \to \infty$, with
$\rho=N/L$ and $n$ held fixed, yields the Poisson distribution,
\begin{equation}
p_n =\frac{\mu^n}{n!}\,\exp(-\mu),
\end{equation}
with $\mu[z]  = \rho R[z]$, the  notation emphasizing that
$\mu$   is  a   {\em  functional}   of  the   $A$-particle  trajectory
$z(t)$.  The Poisson distribution implies that $\mu[z]$ is
the  mean number  of  distinct  $B$ particles  that  have met the
$A$-particle up  to time $t$.  Its derivative, $\dot{\mu}(t)$,  is the
crossing  rate:  $\dot{\mu}(t)\,dt$  is  the probability  that  a  $B$
particle meets the  $A$ particle for the first time  in the time
interval $(t,t+dt)$.

The probability  that the trajectory $z(\tau)$  has survived, in
the original {\em interacting} problem, is simply the probability that
there have been no crossings:
\begin{equation}
p_0[z] = \exp(-\mu[z]).
\end{equation}   
Finally,  the survival  probability  $Q(t)$ is  obtained by  averaging
$\exp(-\mu[z])$  over  all  possible  $A$-particle  trajectories
$z(\tau)$ with the Wiener measure:
\begin{equation}
Q(t) = \frac{\int Dz(t)\,\exp\left[-\frac{1}{4D_A}\int_0^t d\tau
\left(\frac{dz}{d\tau}\right)^2   -  \mu[z]\right]}
{\int Dz(t)\,\exp\left[-\frac{1}{4D_A}\int_0^t                    d\tau
\left(\frac{dz}{d\tau}\right)^2 \right]}\ ,
\label{Q}
\end{equation}
where  the  path-integrals are  restricted  to functions $z(t)$
satisfying $z(0)=0$. The `action functional'
\begin{equation}
S[z] = \frac{1}{4D_A}\int_0^t d\tau\,
\left(\frac{dz}{d\tau}\right)^2 + \mu[z],
\end{equation}
appearing  in the  numerator of  Eq.\ (\ref{Q})  defines  an effective
dynamics for surviving $A$-particle trajectories.

The  functional  $\mu[z]$  can  be  determined  as  follows.  We
calculate, in two ways, the probability density to find a $B$ particle
at the end-point of the  trajectory, $z(t)$, at time $t$. First,
since  the  particles are  treated  as  non-interacting,  and the  $B$
particles  start in  a steady-state  configuration of  uniform density
$\rho$, this probability density is simply $\rho$. It follows that
\begin{equation}
\rho = \int_0^t dt'\,\dot{\mu}(t')\,G(z(t),t|z(t'),t')\ ,
\label{fundamental}
\end{equation}
where, on  the right-hand side, $\dot{\mu}(t')dt'$  is the probability
for a  $B$ particle to  have its first  encounter with the $A$  in the
time interval $(t', t' + dt')$, and
\begin{eqnarray}
G(z(t),t|z(t'),t') &  = & \frac{1}{[4\pi D_B(t-t')]^{1/2}}
\nonumber                 \\                 &&                 \times
\exp\left[-\frac{[z(t)-z(t')]^2}{4D_B(t-t')}\right]
\label{Green}
\end{eqnarray}
is  the  probability density  for  this  $B$-particle to  subsequently
arrive  at $z(t)$ at  time $t$.  Eq.\ (\ref{fundamental})  is an
implicit  equation  for  the  functional $\mu[z]$  (noting  that
$\mu(t=0)=0$, since no $B$ particle  can meet the $A$ particle in zero
time). The continuum version presented here was first derived in ref.\ 
\cite{BMB}. 
The remainder of the paper deals with some of the consequences of the 
`fundamental equation' (\ref{fundamental}).

\section{Exact Asymptotics of $Q(t)$}
\label{Asymptotics}

Before  dealing with the  pre-asymptotic behavior,  which is  the main
focus of this paper, we briefly review  the derivation of the
leading    asymptotic    behavior    first   presented    in    refs.\
\cite{BB,BB2,BMB}.  The argument is  based on deriving upper and lower
bounds  for $Q(t)$  and showing  that these  bounds pinch  for  $t \to
\infty$. 

First consider  the case where  the $A$ particle is  stationary (i.e.\
$D_A=0$),   the  so-called   `target   annihilation  problem',   where
$\vec{z}(t)=0$ for  all $t$.  This problem can  be easily solved  by a
variety of methods \cite{Tachiya,target}. Here we treat it as a  simple  
application of  Eq.\  (\ref{fundamental}).  We put $z=0$ in  
Eq.\ (\ref{Green}), and  call the  corresponding solution of 
(\ref{fundamental}) $\mu_0(t)$.  It satisfies the equation
\begin{equation}
\rho = \int_0^t dt'\,\dot{\mu}_0(t')[4\pi D_B(t-t')]^{-1/2}\ .
\label{mu_0}
\end{equation}
This equation  is readily  solved by taking  the Laplace  transform of
both sides, since  the right-hand side has the  form of a convolution.
Solving  for  $\tilde{\mu}_0(s)$,  and  inverting the  transform,  the
result  is $\mu_0(t)  =  \lambda_1   t^{1/2}$,  as  in
(\ref{BL}), with $\lambda_1  = \rho D_B^{1/2} a_1$ and  $a_1$ given by
(\ref{a}).  
 
It  is intuitively  clear  that  $Q(t)$ for the  system with  $D_A=0$
(stationary  $A$ particle)  provides an  upper bound  on $Q(t)$  for a
system with $D_A>0$. This bound can be proved rigorously \cite{BMB} 
using Eq.\ (\ref{fundamental}). First we write $\mu = \mu_0 + \mu_1$ 
in (\ref{fundamental}), and write, in a more compact notation,
\begin{equation}
G(t_1,t_2) = \frac{1}{[4\pi D_B(t_1-t_2)]^{1/2}}[1 - K(t_1,t_2)],
\end{equation}
where we have taken $t_1 \ge t_2$ without loss of generality, and
\begin{equation}
K(t_1,t_2) = 1 - \exp\{-[z(t_1)-z(t_2)]^2/4D_B(t_1-t_2)\}.
\label{K}
\end{equation}
Eq.\ (\ref{fundamental}) can then be rearranged in the form
\begin{equation}
\int_0^t    \frac{dt'}{(t-t')^{1/2}}\,\dot{\mu}_1(t')    =    \int_0^t
\frac{dt'}{(t-t')^{1/2}}\,\dot{\mu}(t')[1-K(t,t')],
\end{equation}
where  the full  $\dot{\mu}$ appears  on the  right-hand  side. Taking
Laplace transforms of both  sides, solving for $\tilde{\mu}_1(s)$, and
inverting the transform gives
\begin{eqnarray}
\mu_1[z] &=& \frac{1}{\pi}\int_0^t \frac{dt_1}{(t-t_1)^{1/2}}\nonumber \\  
&& \times \int_0^{t_1}\frac{dt_2}{(t_1-t_2)^{1/2}}\,\dot{\mu}(t_2)K(t_1,t_2).
\label{ubound}
\end{eqnarray}
Eq.\ (\ref{ubound}) is an implicit equation for $\mu_1(t)$ because the
full  $\mu =  \mu_0 +  \mu_1$ appears  on the  right-hand  side. Note,
however,  that  $K(t_1,t_2) \ge  0$  and  $\dot{\mu}  \ge 0$  (because
$\mu(t)$ -- the  mean number of different $B$  particles that have met
the  $A$  particle up  to  time $t$  --  is  clearly a  non-decreasing
function).   Therefore   $\mu_1[z]   \ge   0$  for   all   paths
$z(t)$,  with  equality  when  $z(t)=0$ for  all  $t$.  It follows that  
$\mu[\vec{z}] \ge \mu_0(t)$ for  all paths $\vec{z}(t)$, and
\begin{equation}
Q(t) \le Q_U(t) = \exp[-\mu_0(t)]\ .
\label{upper}
\end{equation}
An equivalent result on the lattice, in any space dimension,  was 
obtained in ref.\ \cite{MOBC}. 

This rigorous upper bound for  $Q(t)$ has the same asymptotic behavior
as the rigorous lower bound derived in \cite{BB,BB2}, proving that the
asymptotic  form of  $Q(t)$ is  the same  as for  the `target'problem,
where the $A$ particle is stationary. We sketch the derivation  of the 
lower bound in $d=1$. The results for other dimensions in the  range 
$d \le 2$ can be  found in \cite{BB,BB2}.  To derive the $d=1$ result 
we can calculate the probability for a subset
of  histories for  which the  $A$  particle survives.   These are  the
histories in  which (i) The region $(-l/2,l/2)$  initially contains no
$B$ particles; (ii)  No B-particles enter this region  up to time $t$;
and  (iii)  the $A$  particle  stays within  this  region  up to  time
$t$. The probabilities associated  with (i)-(iii) are $\exp(-\rho l)$,
$\exp[-\mu_0(t)]$,  and (asymptotically) $\sim \exp(-\pi^2D_A  t/l^2)$
respectively. The  product of these probabilities gives,  for any `box
size'  $l$, a  lower  bound  on $Q(t)$.   Optimizing  this bound  with
respect to $l$, at any given time $t$, gives the best lower bound:
\begin{equation}
Q(t) \ge Q_L(t) \sim \exp[-\mu_0(t)- 3(\pi^2\rho^2D_At/4)^{1/3}].
\label{lower}
\end{equation}
Since  $\mu_0(t) \sim  t^{1/2}$  for  $d=1$,  the   second  term  in
(\ref{lower})  is  subdominant  for  large  $t$, and  the  two  bounds
agree. This proves  that the target problem ($D_A=0$)  gives the exact
asymptotic form of $Q(t)$ for any $D_A$. Extensions of this result to
general dimension $d \le 2$ are given in \cite{BB2}.

In  the following  section  we  discuss how  one  might calculate  the
leading pre-asymptotic  correction to $Q(t)$.  

\section{Pre-Asymptotic Corrections}
\subsection{preliminaries}

The starting point  of the calculation is Eq.\  (\ref{Q}), which is an
exact  expression for  the  survival probability,  $Q(t)$,  of an  $A$
particle   with  diffusion  constant   $D_A$  moving   in  a   sea  of
noninteracting  $B$  particles  with  diffusion constant  $D_B$.   The
functional  $\mu[z]$ appearing  in this  equation  is determined
implicitly by the fundamental  equation (\ref{fundamental}). It can be
determined   perturbatively  as   follows.  Writing   $\mu[z]  =
\mu_0[z]  + \mu_1[z]$, where  $\mu_0 =  \lambda_1 t^{1/2}$, we  have  
shown  that $\mu_1[z]$  is  given by  Eq.\ (\ref{ubound})  
where $K(t_1,t_2)$  is  given by  Eq.\ (\ref{K}).  For small diffusion  
constant, $D_A$, of  the $A$ particle,  the typical excursion of the end 
point, $z(t)$, of the  $A$ particle trajectory will be small, and the 
function $K$ can be expanded in powers of $z^2$:
\begin{equation}
K(t_1,t_2)    =    \frac{[z(t_1)-z(t_2)]^2}{4D_B(t_1-t_2)}
-\frac{1}{2}\frac{[z(t_1)-z(t_2)]^4}{[4D_B(t_1-t_2)]^2} + \cdots
\label{Kexp}
\end{equation} 
This  expansion is equivalent,  as we  shall see,  to an  expansion in
powers of $D_A$.
Replacing $K(t_1,t_2)$  in the right-hand side  of Eq.\ (\ref{ubound})
by the first term  in the expansion (\ref{Kexp}), and $\dot{\mu}(t_2)$
by  $\dot{\mu}_0(t_2)$,  where  $\mu_0=\lambda_1 t^{1/2}$,  gives,  to
first order in $D_A$,
\begin{eqnarray}
\mu_1[z]  &  =   & \frac{\lambda_1}{8\pi D_B}
\int_{0}^{t}  \frac{dt_1}{(t-t_1)^{1/2}}   \nonumber  \\  
&& \times \int_{0}^{t_1}\frac{dt_2}{t_{2}^{1/2}(t_1-t_2)^{3/2}}
[z(t_1)-z(t_2)]^2, \nonumber \\ &&
\label{eq:mu1quadrat}
\end{eqnarray}
which  is  a  quadratic  functional  of  the  $A$-particle  trajectory
$z(t)$. Working at the quadratic level will be adequate to first
order in $D_A$.

\subsection{Survival probability}

Using the quadratic expression, Eq. (\ref{eq:mu1quadrat}), for
$\mu_1[z]$,  we  have  the   following  expression  for  the  survival
probability:
\begin{equation}
Q(t) =  \exp(-\lambda_1 t^{1/2})\, \frac{\int  Dz(t) \exp(-S[z])}{\int
Dz(t) \exp(-S_0[z])}
\label{Q1}
\end{equation}
where  both path  integrals  are over  all  trajectories with  initial
points $z(0)=0$,  but with the end points,  $z(t)$, unconstrained. The
`action functionals' $S[z]$ and $S_0[z]$ are given by
\begin{eqnarray}
S[z] &  = & S_0[z] +  \mu_1[z] \\ S_0[z] &  = & \frac{1}{4D_A}\int_0^t
d\tau\,{\dot{z}(\tau)}^2\ .
\label{SS0}
\end{eqnarray} 
Let us  call $Q_1(t)$  the ratio of  path integrals appearing  in Eq.\
(\ref{Q1}),   i.e.\  $Q(t)   =   \exp(-\lambda_1t^{1/2})Q_1(t)$.   The
calculation  of  the  leading-order  contribution  to  the  asymptotic
behavior   of   $Q_1(t)$   can   be  obtained   by   restricting   the
sums-over-paths to those which start and end at $z=0$, i.e.\ paths for
which $z(t)=z(0)=0$. We can see this as follows.

The expression for $Q_1(t)$ can be rewritten in the form
\begin{equation}
Q_1(t)  = \frac{\int_{-\infty}^\infty dx  \int_{z(0)=0}^{z(t)=x} Dz(t)
\exp(-S[z])}  {\int_{-\infty}^\infty  dx \int_{z(0)=0}^{z(t)=x}  Dz(t)
\exp(-S_0[z])}
\label{Q11}
\end{equation}  
Let  $z_c(x,\tau)$  and  $z_{co}(x,\tau)$  be  the  trajectories  that
minimize the functionals $S[z]$  and $S_0[z]$ respectively for boundary
conditions  $z(0)=0$,  $z(t)=x$,  and  write $z(\tau)  =  z_c(x,\tau)+
\tilde{z}(\tau)$   in   $S[z]$    and   $z(\tau)   =   z_{co}(x,\tau)+
\tilde{z}(\tau)$ in $S_0[z]$.  Because both functionals are quadratic,
we obtain immediately
\begin{eqnarray}
Q_1(t)  &  =  &\frac{\int_{-\infty}^\infty dx  \exp\{-S[z_c(x,t)]\}  }
        {\int_{-\infty}^\infty    dx    \exp\{-S_0[z_{co}(x,t)]\}    }
        \nonumber               \\              &&              \times
        \frac{\int_{\tilde{z}(0)=0}^{\tilde{z}(t)=0}      D\tilde{z}(t)
        \exp(-S[\tilde{z}])}{\int_{\tilde{z}(0)=0}^{\tilde{z}(t)=0}
        D\tilde{z}(t) \exp(-S_0[\tilde{z}])}.
\label{Q12}
\end{eqnarray}

In the first factor in (\ref{Q12}), the actions $S$ and $S_0$ are both
proportional   to   $x^2$.   For   the   `free'   action,  $S_0[z]   =
(1/4D_A)\int_0^t  \dot{z}^2  d\tau$, the  path  $z_{co}$  is given  by
$z_{co}(x,\tau)   =   x\tau/t$  and   the   corresponding  action   is
$S_0[z_{co}]  =  x^2/4D_At$.   For  the  action $S[z]$  one  can  show
\cite{BMB} that  $S[z_c] \propto x^2/t^\phi$ with $\phi  \ge 1/4$. The
integrals  over $x$  in the  numerator  and denominator  of the  first
factor  in  (\ref{Q12})  therefore  yield  just  powers  of  $t$.   By
contrast,  the second  factor gives  the exponential  of  a  positive
power of $t$.   This second factor provides the  leading correction to
asymptopia,  and  will  therefore  be  the  focus  of  the  subsequent
development.   To  summarize,   therefore,  we  will  investigate  the
asymptotic behavior of the quantity
\begin{equation}
\hat{Q}_1(t) = \frac{\int_{\tilde{z}(0)=0}^{\tilde{z}(t)=0} D\tilde{z}(t) 
\exp(-S[\tilde{z}])}{\int_{\tilde{z}(0)=0}^{\tilde{z}(t)=0} D\tilde{z}(t) 
\exp(-S_0[\tilde{z}])}.
\label{Q1hat}
\end{equation}
Since the  trajectories $\tilde{z}(\tau)$ vanish at  both $\tau=0$ and
$\tau=t$ we can express them as a Fourier sine series:
\begin{align}
  \tilde{z}(\tau) & =\sum_{n=1}^\infty a_n \sin(n\pi\tau/t)
  \label{eq:sinexp}\\ a_n & =\frac{2}{t}\int_{0}^{t} d\tau 
\tilde{z}(\tau)\sin(n\pi\tau/t).
\label{Fourier}
\end{align}

We substitute the  expansion (\ref{eq:sinexp}) into Eq. (\ref{Q1hat}),
where the functionals $S_0$ and  $S$  are given  by  (\ref{SS0})  and
$\mu_1[z]$ by (\ref{eq:mu1quadrat}). After the rescaling
$a_n \rightarrow 2 [(D_A t)^{1/2}/\pi n] a_n$ we have:
\begin{equation}
\label{eq:psquadratic}
\hat{Q}_1(t)=\int\prod_{n=1}^\infty\Bigl(\frac{da_n}{\sqrt{2\pi}}\Bigr)   
e^{-\frac{1}{2}\sum_n a_{n}^2 -\frac{1}{2}g \sum_{m,n} A_{mn} a_m a_n},
\end{equation}
where the dimensionless coupling constant $g$ is given by
\begin{equation}
g = \frac{D_A}{D_B}\frac{\lambda_1 t^{1/2}}{\pi^3} =
\frac{4\rho}{\pi^{7/2}}\,\frac{D_A}{D_B^{1/2}}\, t^{1/2}
\label{g}
\end{equation}
and the matrix elements $A_{mn}$ are given by
\begin{eqnarray}
A_{mn} & = & \frac{1}{mn}\int_{0}^{1}\frac{dx}{\sqrt{1-x}}
\int_{0}^{x}  \frac{dy}{y^{1/2}(x-y)^{3/2}} \nonumber  \\  && \times
[\sin(m\pi x)-\sin(m\pi y)]\,[\sin(n\pi x)-\sin(n\pi y)].  \nonumber
\\ &&
\label{eq:amatrix}
\end{eqnarray}
We  see  that the  time  dependence  in  the sub-leading  contribution
$\hat{Q}_1(t)$  to $Q(t)$  enters only  through the  coupling constant
$g$.  We note that the prefactor $(4\rho/\pi^{7/2})(D_A/D_B^{1/2})$ is
small  even  for   $D_A=D_B=1$  and  $\rho   =  1$,   where  the
`pre-asymptotic  regime' will  be reached  only for  $t\gg  \pi^7 \sim
3\times  10^3$. At this  timescale, the  leading asymptotic  result is
$Q(t) \sim \exp(-\lambda_1 t^{1/2}) \sim 10^{-30}$. This already gives
strong hints as to why the asymptotic regime is so hard to reach.

A full solution  of the problem, at the  quadratic level, requires the
spectrum of the  matrix $A_{mn}$. At the moment we  lack such an exact
solution, so instead we devise a two-step approach to the problem: (i)
we obtain an analytical lower bound for $\hat{Q}_1(t)$; (ii)
we calculate numerically the spectrum  of the matrix $A$, truncated at
a finite dimension $N$, and we make a finite-size scaling analysis for
the sub-leading correction $\hat{Q}_1(t,N)$ at finite $N$.

\subsection{A Lower Bound on $\hat{Q}_1(t)$}

Consider  Eq.\  (\ref{eq:psquadratic}).   By extracting  the  diagonal
terms,  $(g/2)\sum_n A_{nn}  a_n^2$, from  the quadratic  form  in the
exponential, this equation can be rewritten as
\begin{equation}
\hat{Q}_1(t)=\prod_{n=1}^\infty(1+gA_{nn})^{-1/2}\left\langle
\exp\left[-g\sum_{m<n} A_{mn}a_ma_n\right]\right\rangle
\end{equation}
where the average is performed with the Gaussian weight
\begin{equation}
\exp\left[-\frac{1}{2}\sum_n (1 +g A_{nn}) a_n^2\right].
\end{equation}
Here we  are using  the diagonal  part of the  action, in  the Fourier
basis, as a trial action  functional in a variational calculation. The
variational property  follows from the  convexity inequality, $\langle
\exp x \rangle \ge \exp\langle x \rangle$, which immediately implies
\begin{equation}
\hat{Q}_1(t)\ge \hat{Q}_{1l}(t),
\end{equation}
where
\begin{equation}
\hat{Q}_{1l}(t) = \exp\left[-\frac{1}{2} \sum_n \ln(1+gA_{nn})\right]
\label{Qdiag}
\end{equation}
is a  lower bound for $\hat{Q}_1(t)$.   In the Appendix  we show that,
for large  $n$, $A_{nn}\approx \sqrt{2}\pi^2n^{-3/2}$.  For large $g$,
corresponding  to large  $t$, the  sum on  $n$ is  dominated  by large
values of $n$, of order $n \sim g^{2/3} \sim t^{1/3}$, and the sum can
be replaced by an integral to leading order:
\begin{eqnarray}
\hat{Q}_{1l}(t)   &    \approx   &   \exp\Bigl[-\frac{1}{2}\sum_{n}\ln
(1+g\sqrt{2}\pi^2n^{-3/2})   \Bigr]   \nonumber   \\   &   \approx   &
\exp(-Ag^{2/3})\ , 
\label{33}
\end{eqnarray}
where
\begin{equation}
A = 2^{1/3}\pi^{7/3}3^{-1/2}\ .
\label{A}
\end{equation}
Using Eq.(\ref{g}) for $g$, the final result takes the form
\begin{equation}
\hat{Q}_{1l}(t) \approx \exp\left[-\frac{1}{\sqrt{3}}
\left(32\rho^2\frac{D_A^2}{D_B}t\right)^{1/3}\right]\ .
\label{newlower}
\end{equation}
It is interesting to compare this result with the rigorous lower bound
(\ref{lower}). The latter is valid  for all $D_A$, while the new bound
is valid, in  principle, only for $D_A \ll  D_B$.  In both expressions
the leading  correction to  asymptopia is a  term of order  $(\rho^2 D
t)^{1/3}$  in the  exponential,  where  $D$ has  the  dimensions of  a
diffusion constant.  In the  bound (\ref{lower}) $D=D_A$, while in the
new bound (\ref{newlower}) $D=D_A^2/D_B$.  For small enough $D_A/D_B$,
therefore,  the new bound  is tighter  than the  old, while  for large
$D_A/D_B$ the old  bound is tighter. This confirms  that the new bound
cannot be valid  in general, but only for  small enough $D_A/D_B$. The
two bounds  cross when $D_A/D_B = 3^{9/2}\pi^2/128  \approx 10.82$, so
the new bound is better  for the case $D_A/D_B=1$ used in simulations.
Although the  new bound is  not strictly valid  for $D_A=D_B$ since
the condition $D_A \ll D_B$ is not satisfied, the above considerations
suggest that $D_A \ll 10 D_B$ may be sufficient in practice.

Data  for  $Q(t)$ were  obtained  using  the  algorithm of  Mehra  and
Grassberger \cite{MG}. The system parameters were $D_A=D_B=D=1/2$  and 
$\rho=1/2$. We  plot,  in  Figure 1,  $-\ln  Q(t)/(\rho^2  Dt)^{1/2}$
against  $\ln(\rho^2 Dt)$. The  asymptotic value  of this  quantity is
$a_1  =  4/\sqrt{\pi} \approx  2.257$.   The  upper  and lower  bounds
(\ref{upper}) and (\ref{lower}) are represented by the lower and upper
dashed lines respectively (note the apparent interchange of bounds due
to the  minus sign in the  definition of the ordinate).  The new lower
bound  (\ref{newlower})  is represented  by  the  dotted  line. It  is
clearly a significant  improvement over the old bound,  and appears to
be converging towards the data at late times.

\begin{figure}
\psfig{file=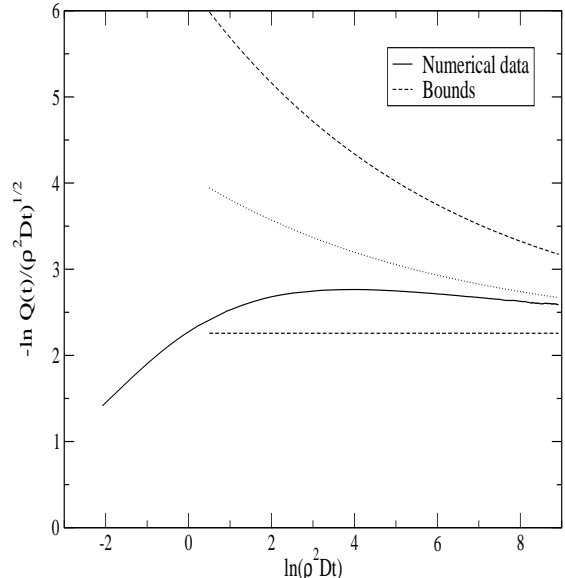,width=8cm,height=7cm,angle=-90,
bbllx=50pt,bblly=50pt,bburx=550pt,bbury=700pt
}
\caption{\label{bounds}Bounds on the quantity $-\ln Q(t)/(\rho^2Dt)^{1/2}$. 
Continuous line: numerical data from ref.\cite{BB2}, with $\rho=1/2$ 
and $D=D_A=D_B=1/2$; dashed lines: rigorous upper and lower bounds; 
dotted line: new lower bound on $Q(t)$ valid for $D_A \ll D_B$.} 
\end{figure}

In the following subsection  we provide a finite-size scaling analysis
which strongly suggests that the lower bound, $\hat{Q}_{1l}(t)$, gives
the  correct asymptotic  form, $\hat{Q}_1(t)  \sim  \exp(-c t^{1/3})$,
although the constant $c$ may differ from (i.e.\ be smaller than) that
appearing in Eq.\ (\ref{newlower}).

\subsection{Exact Diagonalization for finite $N$} 
\label{sec:psnum} 

One can  show that the matrix  elements $A_{mn}$ decay as  a power law
along the rows,  so it is not obvious whether  or not the off-diagonal
matrix elements  make a  significant contribution to  the eigenvalues.
These off-diagonal  terms do  not enter the  calculation of  the lower
bound $\hat{Q}_{1l}$,  so it  is clearly of  interest to  know whether
they qualitatively change the result.
 
As  an analytical insight  is missing  so far,  we study  this problem
numerically  using the  eigenvalues of  finite matrices  $A^N_{mn}$ to
compute  $\hat{Q}_1(g,N)$ of  Eq. (\ref{eq:psquadratic}).   The latter
equation gives
\begin{eqnarray}
\hat{Q}_1   &   =  &   [\det(I   +   gA)]^{-1/2}   \nonumber  \\   &=&
\exp\left(-\frac{1}{2}\sum_{n=0}^{\infty} \ln(1+g\lambda_n)\right)\ ,
\end{eqnarray} 
where $I$ is  the unit matrix and $\lambda_n$  is the $n$th eigenvalue
of   $A$   (where   the   eigenvalues   are   listed   in   decreasing
order). Ideally, in a finite-$N$ analysis we would simply truncate the
sum at $n=N$.  In practice, of course, we do  not know the eigenvalues
explicitly, so instead  we truncate the matrix $A$ to  an $N \times N$
matrix and compute the eigenvalues of the truncated matrix.

In the asymptotic regime we anticipate the finite-size scaling form
\begin{equation}
-\ln[\hat{Q}_1(g,N)]=N^{\alpha}f(g/N^\beta)\ .
\label{FSS1}
\end{equation}
The condition  that $\hat{Q}_1(g,N)$ has  an $N$-independent large-$N$
limit implies  that $f(x) \sim  x^{\alpha/\beta}$ for $  x \rightarrow
0$.

It is instructive to first  consider the diagonal approximation to the
matrix $A$, in which the off-diagonal terms are neglected. This provides
the  lower  bound  $\hat{Q}_{1l}$  given by  Eq.\  (\ref{Qdiag}).  The
corresponding  finite-size quantity,  $\hat{Q}_{1l}(g,N)$ is  given by
$\hat{Q}_{1l}(g,N) =  \exp[-(1/2)\sum_{n=1}^N \ln(1+gA_{nn})$, leading
to
\begin{equation}
-\ln[\hat{Q}_{1l}(g,N)] = \frac{1}{2} \sum_{n=1}^N \ln(1+gA_{nn})\ .
\end{equation} 
For  large  $n$,  we  show  in  the Appendix  that  $A_{nn}$  has  the
asymptotic  form $A_{nn}  \sim n^{-3/2}$,  so the  finite-size scaling
variable is $g/N^{3/2}$, i.e.\  $\beta=3/2$ in (\ref{FSS1}) within the
diagonal approximation.  In the limit  $g \to \infty$, $N  \to \infty$
with $g/N^{3/2}$  held fixed, one can  replace the sum over  $n$ by an
integral and  one readily recovers the scaling  form (\ref{FSS1}) with
$\alpha=1$,  $\beta=3/2$. It  is also  easy to  show that  the scaling
function  $f(x)$  in  Eq.\   (\ref{FSS1})  has,  within  the  diagonal
approximation, the limiting  forms $f(x) \sim x^{2/3}$ for  $x \ll 1$,
as expected, while $f(x) \sim \ln(x)$ for $x \gg 1$.

Our aim is  to investigate whether the same  general scaling structure
(\ref{FSS1}), with the same exponents $\alpha=1$ and $\beta=3/2$, holds
when the  full matrix $A$ is  used instead of just  its diagonal part.
If  the matrix  $A$ is  simply truncated  to an  $N \times  N$ matrix,
however,   and  the   eigenvalues  determined,   the   corrections  to
finite-size  scaling are found  to be  large. We  attribute this  to a
large influence of  the missing matrix elements $A_{mn}$,  with $m$ or
$n$ greater  than $N$,  on the eigenvalues  $\lambda_n$ when  $n$ gets
close to $N$. Instead, therefore,  we adopt a more refined approach in
which the matrix is truncated to  size $2N$, but we use only the first
$N$ eigenvalues to compute $\hat{Q}_{1}(g,N)$. Furthermore, instead of
computing   $-\ln\hat{Q}_{1}(g,N)  =   (1/2)\sum_{n=1}^N   \ln  (1   +
g\lambda_n)$, we compute the derivative
\begin{eqnarray}
H(g,N) &  = &  -2 \frac{d}{dg}\ln\hat{Q}_{1}(g,N) \nonumber  \\ &  = &
\sum_{n=1}^N \lambda_n/(1 + g\lambda_n)\ ,
\label{H}
\end{eqnarray}
since we find  that corrections to finite-$N$ scaling  are smaller for
this quantity.

The finite-size scaling form  for $H(g,N)$ follows from its definition
and Eq.\ (\ref{FSS1}):
\begin{equation}
H(g,N) = N^{\alpha - \beta} h(g/N^{\beta})\ ,
\end{equation}
where $h(x)=2\, df/dx$. On the  basis of the diagonal  approximation we
expect $\alpha = 1$, $\beta =3/2$,  and $h(x) \to (4A/3)x^{-1/3}$ for 
$x \ll 1$, where $A$ is given by Eq.\ (\ref{A}), and, from Eq.\ (\ref{H}),  
$h(x)  \to  x^{-1}$  for   $x  \gg   1$.  We   therefore  plot
$N^{1/2}H(g,N)$ against the scaling variable $x=g/N^{3/2}$.

\begin{figure}
\psfig{file=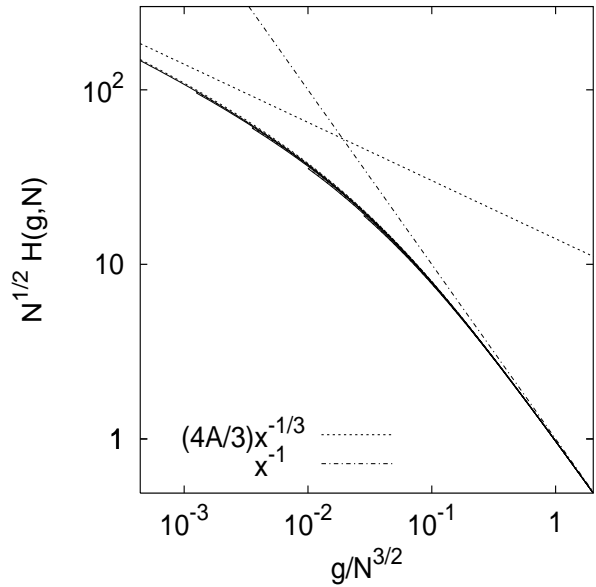,width=15cm,height=8cm,
bbllx=50pt,bblly=50pt,bburx=554pt,bbury=300pt}
\caption{\label{fig:psscaling}Scaling plot for $H(g,N) = -2
(d/dg)\ln[\hat{Q}_1(g,N)]$ ($N=50,100,200,400,800,1600$), with scaling
variable $x=g/N^{3/2}$.  Continuous curves -- full matrix; broken
curves -- diagonal matrix.  The straight lines are the asymptotes for
large and small scaling variable -- see text.}
\end{figure}

Fig.\  \ref{fig:psscaling}  shows  that  a good  scaling  collapse  is
obtained  with $\alpha=1$  and $\beta=3/2$.  This result  implies that
$-\ln\hat{Q}_1 \propto g^{2/3} \propto t^{1/3}$, i.e.\ the nondiagonal
elements of the matrix $A$  do not qualitatively change the scaling of
the  pre-asymptotic  correction   to  the  survival  probability.  The
straight lines in the Figure are the asymptotes dicussed above, with 
slopes $-1/3$ and $-1$  for small and  large scaling variable respectively. 
The main  curve actually contains both the full matrix  (continuous lines) 
and  the diagonal  approximation (symbols), which lie almost on top of 
each other. 

A very  small deviation from perfect  scaling is evident  in the data.
In fact, a slightly better fit is obtained with $\beta=1.4$ instead of
1.5.  However,  the same  numerical  scaling analysis  applied to  the
diagonal  matrix, where  $\alpha=1$  and $\beta=3/2$  are exact,  also
yields,   for  the   range   of   $N$  and   $g$   covered  by   Fig.\
(\ref{fig:psscaling}),  an  apparent  best  fit  with  $\beta  \approx
1.4$. We conclude that, for the full matrix, $\beta=3/2$ is consistent
with the  data within the  numerical precision. We  conclude, finally,
that the  form of  the leading correction  to asymptopia given  by the
lower  bound (\ref{33})  is correct.  Even the  amplitude seems  to be
given rather accurately by the lower bound.

\section{Beyond the quadratic action}

In  the  previous sections  we  have  used  the functional  $\mu_1[z]$
correct  to  first order  in  the  ratio  $D_A/D_B$ of  the  diffusion
constant of the particle to that  of the traps. To this leading order,
the  parameter which  controls the  asymptotic behaviour  is $g\propto
D_A/\sqrt{D_B}$. We obtained the  following asymptotic lower bound for
the survival probability, $Q(t)$, of the $A$-particle in $d=1$:
\begin{equation}
Q(t) \ge \exp[-\lambda_1 t^{1/2} - Ag^{2/3} + \cdots]
\end{equation}
where $g=(D_A/D_B)\lambda_1 t^{1/2}/\pi^3$ and $A$ is a constant. This
suggests,  as  discussed  in  the  preceding section,  that  at  large
$D_A/D_B$ the new  lower bound can cross the  lower bound deduced from
an ``excluded  volume'' kind of argument \cite{BB,BB2},  given by Eq.\
(\ref{lower}).

To  escape  from this  apparent  contradiction,  one  has to  consider
higher-order (in $D_A/D_B$) terms in  the action. To go to next order,
we start from  Eq.\ (\ref{ubound}), and expand the  right-hand side up
to order  $z^4$.  After a tedious but  straightforward calculation, we
obtain a  result for  the term  order $z^4$ in  the form  of a  sum of
(multiple) integrals.  Inserting the Fourier expansion (\ref{Fourier})
and rescaling the expansion coefficients  $a_n$ as before to make them 
dimensionless, we can cast the action in the dimensionless form
\begin{equation}
S = S_0+gS_1+g\frac{D_A}{D_B}S_{2} + O([D_A/D_B]^2),
\label{S2}
\end{equation}
where
\begin{align}
  S_0&=\frac{1}{2}\sum_{n}a_{n}^2\\
  S_1&=\frac{1}{2}\sum_{m,n}A_{mn}a_{m}a_{n}\\
  S_{2}&=\frac{1}{4}\sum_{m,n,p,q}A_{mnpq} a_{m}a_{n}a_{p}a_{q}
\end{align}
and  the   vertices  $A_{mn}$,  $A_{mnpq}$  are   pure  numbers.  Eq.\
(\ref{S2})  enables us  to extend  the calculation  of $\hat{Q}_1(t)$,
Eq.\  (\ref{eq:psquadratic}),  to one  higher  order  in $D_A/D_B$  by
replacing the exponential in Eq.\ (\ref{eq:psquadratic}) by $\exp(-S)$
with $S$ given by Eq.\ (\ref{S2}).

While we have  not carried out this program  explicitly, the structure
of the expansion (\ref{S2})  shows that the quadratic approximation is
valid for  small $D_A/D_B$.   What is the  effect of the  higher order
terms?  For  the case  $D_B=0$  (where  the  traps are  immobile)  the
asymptotic behavior has, in  $d=1$, the known functional form $\exp(-c
t^{1/3})$, where $c \propto \rho D_A^{1/3}$. It is plausible to assume
that higher-order corrections change the coefficient $c$ but leave the
exponent  unaltered.  Then,  if  the  above  series  in  $D_A/D_B$  is
convergent, we obtain the asymptotic behavior of $Q(t)$, including the
leading pre-asymptotic correction, in the form
\begin{equation}
Q(t)   \sim  \exp\left[-\frac{4}{\sqrt{\pi}}(\rho^2  D_B   t)^{1/2}  -
(\rho^2 D_A t)^{1/3} F\left(\frac{D_A}{D_B}\right)\right]\ .
\label{conj}
\end{equation}
The function $F(x)$  has the limiting behavior $F(x)  \to A_1 x^{1/3}$
for  $x  \to  0$,  with  $A_1 \le  (32)^{1/3}/\sqrt{3}$  from the lower 
bound (\ref{newlower}),  while $F(x)  \to  A_2$ for  $x  \to \infty$,  
where $A_2=3/2^{2/3}$ to  match the known asymptotic  behavior with 
immobile traps --  the so-called Donsker-Varadhan problem \cite{DV}. 
This  latter  limit  agrees  with  Eq.\  (\ref{lower})  (where
$\mu_0=0$ for  immobile traps). The  reason is simple. In  $d=1$, only
the two immobile traps immediately  to the left and right respectively
of the  $A$ particle play any  role. Let the distances  of these traps
from  the $A$-particle  be $x$  and $y$  at $t=0$.  Then  the particle
survives as long as it stays within the box of size $(x+y)$ defined by
these two traps, so its asymptotic survival probability, averaged over
$x$  and  $y$,  is  given   by  $Q(t)  \sim  \rho^2  \int_0^\infty  dx
\int_0^\infty dy  \exp[-\rho(x+y) - \pi^2  D_A t/(x+y)^2]$. Evaluating
the integral by steepest descents for $t \to \infty$ one recovers Eq.\
(\ref{lower}) (with $\mu_0=0$).

\section{Summary and Conclusion}
In this paper we have addressed the issue of the leading correction to
asymptopia in the long-time dynamics  of the trapping reaction, $A + B
\to B$.  The main results are a new rigorous lower bound, valid in the
limit $D_A \ll D_B$, on  the survival probability of the $A$-particle,
and  a numerical  analysis, based  on finite-size  scaling, indicating
that the  new lower bound contains  the correct analytic  form for the
correction to asymptopia, namely a $t^{1/3}$ correction to the leading
$t^{1/2}$  term in the  exponential. The new bound agrees well, 
at late times, with numerical data, as shown in Figure 1. We  also 
presented  a conjecture, Eq.\ (\ref{conj}),  for the form  of the 
correction to  asymptopia for general $D_A/D_B$,  incorporating both  
extreme limits, $D_A  \ll D_B$ and $D_A \gg D_B$.

Open   questions  to  be   addressed  in   future  work   include  the
generalization  of   the  work   presented  here  to   higher  spatial
dimensions. 
Another interesting question, addressed briefly in \cite{BMB}, is the 
scaling with time of the  fluctuations of  surviving  trajectories.  
These  can be  studied
starting  from  Eq.\  (\ref{eq:mu1quadrat}),  using the  same  Fourier
decomposition of the  trajectory as employed in this  work.  One finds
that the fluctuations are  subdiffusive as suggested in \cite{BMB} and
observed  in  numerical   simulations  \cite{MG}.   Details  of  these
calculations will be presented in a separate publication.
  
\section*{Acknowledgements}

We thank Richard Blythe for discussions. LA acknowledges support from 
the European Community Marie Curie Fellowship scheme under contract 
No. HPMF-CT-2002-01910. 

\appendix

\section{Diagonal elements of the matrix $A$}\label{sec:amat}

The matrix element $A_{mn}$ is given by the double integral
\begin{eqnarray}\label{eq:amninit}
  I_{mn} &=& mnA_{mn}=\int_{0}^{1}\frac{dx}{\sqrt{1-x}}\int_{0}^{x}
  \frac{dy}{\sqrt{y}(x-y)^{3/2}} \nonumber \\ 
&& \times (\sin(n\pi x)-\sin(n\pi y)) (\sin(m\pi x)-\sin(m\pi y)) 
\nonumber \\
&&  
\end{eqnarray}
which by variable substitutions $z=x-y$, $s=y$ and standard
manipulations of the trigonometric functions can be reduced 
to a sum of one-dimensional integrals
\begin{eqnarray}\label{eq:Imn1d}
  I_{mn}&=& 2\pi\cos\frac{\pi(m-n)}{2} \int_{0}^{1}dz
  \frac{1}{z^{3/2}}\sin\frac{n\pi z}{2} \sin \frac{m\pi
  z}{2} \nonumber \\
&& \times J_0\left(\frac{\pi}{2}(m-n)(1-z)\right) \nonumber \\ 
&& +2\pi\cos \frac{\pi(m+n)}{2}
  \int_{0}^{1}dz \frac{1}{z^{3/2}}\sin\frac{n\pi z}{2} 
\sin \frac{m\pi z}{2} \nonumber \\ 
&& \times J_0\left(\frac{\pi}{2}(m+n)(1-z)\right),
\end{eqnarray}
where $J_0(z)$ is the Bessel function of the first kind. The fact that
$I_{mn}=0$  if $m+n=2p+1$ can  be obtained  from the  initial formula,
Eq.\ (\ref{eq:amninit}),  if one notices that the  kernel is symmetric
at reflection about the line $y=1-x$.

For the diagonal elements, $m=n$, the leading large-$n$ contribution 
comes from the first integral
\begin{eqnarray}
  2\pi\int_{0}^{1}dz \frac{1}{z^{3/2}}\sin^2\frac{n\pi z}{2} 
& \approx & 2\pi \sqrt{n}\int_{0}^{\infty} \frac{1}{z^{3/2}}
\sin^2\frac{\pi z}{2} \nonumber \\
& = & \sqrt{2}\pi^2 n^{1/2}
\end{eqnarray}
correct to leading order for large $n$. It follows that
\begin{equation}
A_{nn} \to \sqrt{2}\pi^2 n^{-3/2}
\end{equation}
for $n \to \infty$. 

The calculation of  the asymptotic behaviour of $A_{mn}$  at large $n$
with $m$  fixed and large  is more elaborate.  Since it is  not needed
here, we will defer it to a future publication.

\end{document}